\documentclass[12pt]{article}
\usepackage{fullpage}
\usepackage{setspace}
\usepackage{lscape}
\usepackage{amsmath}
\usepackage{amsthm}
\usepackage{acronym}
\usepackage{graphicx}
\usepackage{amssymb}
\usepackage{bm}
\usepackage{mathtools}
\usepackage{booktabs}
\usepackage{multirow}
\usepackage{hyperref}
\usepackage{setspace}
\usepackage{natbib}
\usepackage{color}
\usepackage{caption}
\usepackage{amsfonts}
\usepackage{bm}
\usepackage{mathtools}
\usepackage{booktabs}
\usepackage{mathrsfs}
\usepackage{mathptmx}
\usepackage{multirow}
\usepackage{mathrsfs}
\usepackage{subcaption}
\usepackage{latexsym}
\usepackage{nicefrac}
\usepackage{verbatim}
\usepackage{lineno}

\everydisplay{\abovedisplayshortskip\belowdisplayshortskip}
\makeatletter \oddsidemargin  -.1in \evensidemargin -.1in

\theoremstyle{plain}
\newtheorem*{Lemma*}{Lemma}
\title{Inequality Restricted Estimator for Gamma Regression: Bayesian approach as a solution to the Multicollinearity}
\author{Solmaz Seifollahi$^1$, Hossein Bevrani$^1$ and Kaniav Kamary$^2$\\
\small {$^1$ Department of Statistics, Faculty of Mathematical Science, University of Tabriz, Tabriz, Iran}\\
\small {$^2$ Fédération Mathématique, CentraleSup\'elec, Universit\'e Paris-Saclay, Gif-sur-Yvette, France}\\
\small{s.seifollahi@tabrizu.ac.ir; bevrani@tabrizu.ac.ir; kaniav.kamary@centralesupelec.fr}
}

\date{}

\begin{document}

\maketitle

\noindent{\bf {\em Abstract:}}
In this paper, we consider the multicollinearity problem in the gamma regression model when model parameters are linearly restricted.
 The linear restrictions are available from prior information to ensure the validity of scientific theories or structural consistency based on physical phenomena. In order to make relevant statistical inference for a model any available knowledge and prior information on the model parameters should be taken into account.  This paper proposes therefore an algorithm to acquire Bayesian estimator for the parameters of a gamma regression model subjected to some linear inequality restrictions.
We then show that the proposed estimator outperforms the ordinary estimators such as the maximum likelihood and ridge estimators in term of pertinence and accuracy through Monte Carlo simulations and application to a real dataset.

\vskip 3mm
\noindent {\bf {\em  Keywords and phrases:}} Bayesian Inference, Gamma Regression Model, Linear Inequality Restrictions, Prior Modelling.
\vskip 3mm
\noindent {\bf {\em {\color{blue} MSC2020 subject classifications:}}} 62F15; 62J12.
\vskip 6mm

\section{Introduction}
The gamma regression model is a well-known method for assessing data in medical science, health-care economics, and automotive insurance claims.
It gives an effective and flexible way to model and forecast the response variable when we have a positively skewed continuous data following a gamma distribution.
In general, for the generalized linear regression models, the full independence of the covariates is often considered as an assumption to simplify the estimation procedure.
However, this assumption does not always hold in practice and leads to the problem of multicollinearity.
One of the consequences of the multicollinearity is that the maximum likelihood estimation of the coefficients of the gamma regression model becomes unreliable with high variance and consequently, lower statistical significance \citep{Dunder,Malehi}.
To overcome the multicollinearity issu, various methods have been proposed as remedial methods and for instant, \cite{Segerstedt} considered the ridge estimator in the generalized linear models.
Furthermore, in some cases, we must impose some linear inequality constraints on regression parameters based on prior knowledge we have on the model parameters.
As an example, in hyper-spectral imaging, due to physical considerations, the coefficient parameters should be non-negative \cite{Manolakis}.
Another examples may be found in the fields of astronomy and zoology \citep{Wang}, in the field of geodesy \citep{Zhu}, and in the fields of economics and biology \citep{Silvapulle}.
The linear inequality restrictions in linear regression models have been widely investigated to find the least square estimator and its properties such as bias, the mean square error and the efficiency over inequality restrictions (see \cite{Judge}, \cite{Liew}, \cite{Lovell}, \cite{Escobar1,Escobar2}, and \cite{Ohtani}).
Recently, in the area of big data, it has been proved that the incorporation of non-negative restrictions provides sparsity without using any regularization in linear regression models \citep{Meinshausen,Slawski} and also in generalized linear models \citep{Koike}.
Obviously, ignoring this type of information can impact the model estimation procedure, and decrease the accuracy of predictions.
Bayesian method takes into account all prior information available about the model parameter and combine it with evidence from information contained in a sample to guide the statistical inference process. Hence, Bayesian models always provide an easy way to trickle such prior information based on the linear inequality restrictions in the model estimation process. \cite{Geweke1, Geweke2}, \cite{Neal}, \cite{Neelon} and recently \cite{Veiga} are some examples of the research on the Bayesian inference of the linear regression models subject to linear inequality restrictions. 
In general, the majority of the literature on Bayesian inference with linear inequality restrictions on the parameters has concentrated on the linear regression models. However, such restrictions are common to occur in many studies where the generalized linear models may be applicable. For generalized linear models, \cite{Ghosal} introduced an algorithm to obtain the Bayesian estimation based on the linear inequality restrictions under some conditions that are not always satisfied. For instance, the gamma regression models with the log link function do not satisfy the \cite{Ghosal}'s conditions.
In this paper, we consider a Bayesian solution to make statistical inference on the gamma regression model in the case where the model parameters undergo linear inequality constraints and multicollinearity. 

The structure of the paper is as follows : We describe the gamma regression model and define the maximum likelihood estimator of the regression parameters in Section \ref{sec2}. In Section \ref{sec3}, we introduce the Bayesian estimation method for the model subject to the linear inequality constraints. Then, we compare the performance of our proposed Bayesian estimator to existing methods
using two simulation studies in
Section \ref{sec4}. Section \ref{real_app} contains a real-life dataset analysis and finally, in
Section \ref{conclusion}, we conclude the paper.

\section{Gamma regression model and MLE} \label{sec2}
Gamma regression models are applied to model the response variables with positively skewed distribution similar to the gamma distribution.
Let $y_1, y_2, \cdots, y_n$ be a random sample of the response variable $y$. The commun forme of the gamma probability distribution function is defined as follows :
\begin{equation}\label{dens:Y}
f(y_i| \tau, \delta_i) = \dfrac{1}{\Gamma(\tau) \delta_i^{\tau} } y_i^{\tau-1} \exp\{-\dfrac{y_i}{\delta_i}\}; \qquad y_i \geq 0 ~\text{for}~ i=1, \ldots, n,
\end{equation}
with
\begin{align}\label{mean:y}
  \mathbb{E} (Y_i) & = \tau \delta_i \\
  \mathbb{V} (Y_i) & = \tau \delta_i^2; \qquad i: 1, 2, \cdots, n
\end{align}
where $\tau >0$ is the shape parameter, $\delta_i>0$ is the scale parameter and $\Gamma(.)$ is the gamma function.
In this paper, we consider the following parametrization for the gamma distribution introduced by \cite{Amin} when $\tau= \dfrac{1}{\zeta}$ and $\delta_i= \mu_i \zeta$ :
\begin{equation}\label{rep:pdf}
  f(y_i| \zeta, \mu_i) = \dfrac{1}{\Gamma(\dfrac{1}{\zeta}) (\zeta \mu_i)^{1/\zeta}} y_i^{1/\zeta-1} \exp \{-\dfrac{y_i}{\mu_i \zeta}\}; \qquad y \geq 0,
\end{equation}
where $\tau= \dfrac{1}{\zeta}$ and $\delta_i= \mu_i \zeta$. $\zeta= 1/\tau$ is called the precision parameter and $\mu_i= \zeta \delta_i; i: 1, 2, \cdots, n$ is the mean of $y_i$.
By supposing that $\zeta$ is a priori known, we then use the log function $
\log(\mu_i)= X_i^T \pmb{\beta}$
~so as to link the response variable to the covariate variable $X_i= (x_{i1}, x_{i2}, \cdots, x_{ip})^T; i= 1, 2, \cdots, n$. The matrix $X=[X_1, X_2, \cdots, X_p]^T$ is the design matrix of order $n \times p$ where $n$ is the sample size and $p$ is the number of the covariates in the model and $ \pmb{\beta}= (\beta_1, \beta_2, \cdots, \beta_p)^T$ is the vector of the model coefficients.


The likelihood function based on the forme \eqref{rep:pdf} is then given by
\begin{equation}\label{lik:func}
  \ell (\pmb{\beta}| Y, X, \zeta)= \bigg[\dfrac{e^{[-(1/\zeta) \log(\zeta)]}}{\Gamma(1/\zeta)} \bigg]^n \prod_{i=1}^{n} y_i^{1/\zeta-1} \exp\bigg\{- \dfrac{1}{\zeta} \sum_{i=1}^{n} \bigg[y_i e^{-X_i^T \pmb{\beta}}+ X_i^T \pmb{\beta}\bigg] \bigg\}
\end{equation}

A commun classical method to estimate the model parameters is the maximum likelihood estimation when the following log-likelihood function is maximized with respect to the parameter vector $\pmb{\beta}$ :
\begin{equation}\label{log:lik}
  \log(\ell(\pmb{\beta}| Y, X, \zeta))= -\dfrac{1}{\zeta} \sum_{i=1}^{n} \bigg[y_i e^{-X_i^T \pmb{\beta}}+ X_i^T \pmb{\beta}\bigg]+ \dfrac{1-\zeta}{\zeta}\sum_{i=1}^{n} \log(y_i)- \dfrac{n}{\zeta} \log (\zeta)-n \log \bigg(\Gamma(\dfrac{1}{\zeta})\bigg).
\end{equation}
 The derivative of the log-likelihood function with respect to $\pmb{\beta}$, is then as :
\begin{equation}\label{score:fun}
  \dfrac{\partial}{\partial \pmb{\beta}}  \log(\ell(\pmb{\beta}| Y, X, \zeta))= -\dfrac{1}{\zeta} \sum_{i=1}^{n} \left[y_i e^{-X_i^T \pmb{\beta}}-1\right]X_i
\end{equation}
The derivative function \eqref{score:fun} is a non-linear function in $\pmb{\beta}$ and thus, following the MLE process  presented by \cite{Hardin}, $\pmb{\beta}$ can be estimated by applying the Fisher scoring method whose iterations can be obtained as following 
\begin{equation}\label{IRSE}
  \hat{\pmb{\beta}}^{(t+1)}= \hat{\pmb{\beta}}^{(t)}+ \bigg[ \left[I^{-1}(\pmb{\beta})\right] \dfrac{\partial}{\partial \pmb{\beta}}  \log(\ell(\pmb{\beta}| Y, X, \zeta)) \bigg]_{ \pmb{\beta}=\hat{\pmb{\beta}}^{(t)} }
\end{equation}
in which $
  I(\pmb{\beta})$
~is the Fisher Information matrix.
The iteration process will be carried out until the convergence of the resulting values is attained. The maximum likelihood estimation is then computed  by applying the following iterative re-weighted least square algorithm : 
\begin{equation}\label{MLE}
  \hat{\pmb{\beta}}_{MLE}= \left(X^T \left[\text{diag}(\hat{\mu_i}^2)\right]X\right)^{-1} X^T \left[\text{diag}(\hat{\mu_i}^2)\right] \pmb{M},
\end{equation}
where $\pmb{M}$ is an $n$-length vector with $i$-th element defined by the adjusted response variable $X_i^T \hat{\pmb{\beta}}+ \dfrac{y_i-\hat{\mu_i}}{\hat{\mu_i}^2}$. 
The accuracy of the maximum likelihood estimator may be negatively affected due to the ill-conditioning and the multicollinearity issue of the design matrix \citep{Segerstedt,Mackinnon}. 
 To overcome the multicollinearity issue, \cite{Amin}  used gamma ridge estimator defined as follows :
 \begin{equation}\label{ridge}
 \hat{\pmb{\beta}}_{GRE}= \left(X^T  \left[\text{diag}(\hat{\mu_i}^2)\right]X+ \text{diag}(k)\right)^{-1} X^T  \left[\text{diag}(\hat{\mu_i}^2)\right]X \hat{\pmb{\beta}}_{MLE},
 \end{equation}
where the penalty parameter, $k$, is unknown and must be estimated.

\section{Bayesian inference}\label{sec3}
Suppose that the gamma regression parameter $\pmb{\beta}$ is bounded as following :
\begin{equation}\label{H0}
  \pmb{R}\pmb{\beta} \leq \pmb{r},
\end{equation}
where $\pmb{R}$ and $\pmb{r}$ are respectively a pre-specified $m \times p $-matrix and an $m$-length vector. We assume that the restriction number $m$ can be more than $p$ and the sub-space created by the restrictions in \eqref{H0} is not empty.
 

We assign a truncated multivariate normal distribution as the prior to the parameter $\pmb{\beta}$
\begin{equation}\label{TMVN}
  \pmb{\beta} \sim \mathcal{TN}_p (\pmb{\mu}_{\pmb{\beta}}, \pmb{\Sigma}_{\pmb{\beta}}, \pmb{R}, \pmb{r}),
\end{equation}
with the following probability density function :
\begin{equation}\label{DTVM}
  \pi(\pmb{\beta})= \dfrac{\exp \left\{-\frac{1}{2}(\pmb{\beta}-\pmb{\mu}_{\pmb{\beta}})^T\pmb{\Sigma}_{\pmb{\beta}}^{-1}(\pmb{\beta}-\pmb{\mu}_{\pmb{\beta}})\right\}}{\int_{\pmb{R}\pmb{\beta} \leq \pmb{r}}\exp \left\{-\frac{1}{2}(\pmb{\beta}-\pmb{\mu}_{\pmb{\beta}})^T\pmb{\Sigma}_{\pmb{\beta}}^{-1}(\pmb{\beta}-\pmb{\mu}_{\pmb{\beta}})\right\} d\pmb{\beta}}\mathbb{I}(\pmb{R}\pmb{\beta} \leq \pmb{r}).
\end{equation}
where $\mathbb{I}$ denotes the indicator function and the hyperparameters $\pmb{\mu}_{\pmb{\beta}}, \pmb{\Sigma}_{\pmb{\beta}}$ are supposed to be known.

The posterior distribution of $\pmb{\beta}$ is then given by :
\begin{align}\label{post}
  \pi(\pmb{\beta}|Y, X, \gamma)&
 & \propto \exp\bigg\{- \dfrac{1}{\zeta} \sum_{i=1}^{n} \bigg[y_i e^{-X_i^T \pmb{\beta}}+ X_i^T \pmb{\beta}\bigg] -  \dfrac{1}{2} (\pmb{\beta}-\pmb{\mu}_{\pmb{\beta}})^T\pmb{\Sigma}_{\pmb{\beta}}^{-1}(\pmb{\beta}-\pmb{\mu}_{\pmb{\beta}})\bigg\}\mathbb{I}(\pmb{R}\pmb{\beta} \leq \pmb{r})
\end{align}
 While the posterior distribution does not have a closed form,
we use the Metropolis-Hastings algorithm to derive the Bayesian estimation of the gamma regression coefficients (BEGRC).
We then consider the following random walk truncated normal distribution
\begin{equation}\label{prop:dist}
  \pmb{\beta} \sim \mathcal{TN}_p (\pmb{\beta}^{(t-1)}, \pmb{\Sigma}_{pro}, \pmb{R}, \pmb{r}).
\end{equation}
as the proposal distribution of the algorithm with $\pmb{\Sigma}_{pro}$ constant.

Many algorithms have been proposed to generate samples from truncated multivariate normal distribution subject to linear inequality restrictions such as, \cite{Geweke1, Geweke2}, \cite{Ghosh}, \cite{Rod}, \cite{Pakman}, \cite{Lan}, \cite{Cong}, and most of them are practicable when $m<p$.
We use the sampling method proposed by \cite{Li}, in which samples are generated through a series of Gibbs cycles, each of which amounts to sample from the univariate truncated normal distribution. This procedure can be accomplished through the efficient specialized rejection sampling depending on the type of the constraint.


\section{Simulation Study}\label{sec4}
In this section, the performance of the proposed estimator is illustrated by designing some simulation studies.\\
{\bf Random Data Generation : }
We consider a gamma regression model with four covariates such that
\begin{align}\label{sim:mdl}
  \eta_i & = X_{i1} \beta_1+ X_{i2} \beta_2+X_{i3} \beta_3+X_{i4} \beta_4, \\
  \mu_i & = \exp\{\eta_i\}; \qquad i: 1, 2, \cdots, n
\end{align}
To build the experiment, we first generate the predictors from
\begin{equation*}
  x_{ij}= (1-\rho^2)^{1/2} w_{ij}+ \rho w_{ik}; \qquad i, j: 1, 2, \cdots, n \quad \text{and} \quad k=p+1,
\end{equation*}
where $w_{ij}$ is generated from the standard normal distribution. The parameter $\rho$ controls the intensity of inter-correlation among the covariates.
We consider different values for $\rho$ such as $0.8, 0.90, 0.95$ and $0.99$ and the exact values of the regression coefficients in this experiment are chosen as $\pmb{\beta}= (1,1,1,1)^T$.
To investigate the effect of the sample size on the performance of BEGRC, we do data analysis for samples of sizes  $25, 50, 100$ and $200$.
Finally, the response variables $y_i$ are generated from $\mathcal{G}(1/\zeta, \mu_i \zeta)$ where $\zeta$ is once set to $0.25$ and then to $0.5$.

{\bf Prior modelling : }For the prior specification, we consider $\beta_j\geq 0.8$ when $j=1,\ldots,4$ and 
\begin{equation}\label{Hyper:par}
  \pmb{\mu}_{\pmb{\beta}}= \pmb{0}\qquad \text{and} \qquad \pmb{\Sigma}_{\pmb{\beta}}= (X^T X)^{-1}.
\end{equation}
For the proposal distribution in the MCMC algorithm, the covariance matrix is specified as the inverse function of the Fisher information matrix
\begin{equation}\label{Sig:prop}
  \pmb{\Sigma}_{pro}=  \dfrac{1}{\zeta} \left(X^T \left[\text{diag}(\hat{\mu_i}^2)\right] X\right)^{-1}.
\end{equation}

For each dataset, we compute the BEGRC, the maximum likelihood estimation (MLE) and the ridge estimator by using the expression \eqref{ridge} when the penalty parameter (defined by \cite{Amin}) is once as 
 \begin{equation*}
k_1 =  \dfrac{\lambda_{min} \zeta}{\alpha_{min}^2}
\end{equation*}
and then as
\begin{equation}\label{ridge.tun}
k_2  = max\bigg( \dfrac{\lambda_{max} }{(n-p)\zeta +\lambda_i\alpha_{i}^2} \bigg).
\end{equation}
Note that if $\Lambda$ is the matrix whose columns are the eigenvectors of $X^T\left[\text{diag}(\hat{\mu_i}^2)\right]X$, then in formulas above, $\lambda=(\lambda_1, \ldots, \lambda_p)^T$ is the eigenvalues of the matrix $X^T\left[\text{diag}(\hat{\mu_i}^2)\right]X$, and
$$\alpha=(\alpha_1,\ldots, \alpha_p)^T=\Lambda \hat{\pmb{\beta}}_{MLE}, \quad\alpha_{\min}^2= \underset{1\leq j\leq p}{\min}(\alpha_j^2), \quad \lambda_{\min}=\underset{1\leq j\leq p}{\min} (\lambda_j)~ \text{and}~\lambda_{\max}=\underset{1\leq j\leq p}{\max} (\lambda_j).$$

We also consider the case where the restriction \eqref{H0} is neglected in order to evaluate the impact of ignoring this information on the posterior estimates and provide a Bayesian inference on the model parameters by considering the following prior distribution 

\begin{equation}\label{MVN}
  \pmb{\beta} \sim \mathcal{N}_p (\pmb{\mu}_{\pmb{\beta}}, \pmb{\Sigma}_{\pmb{\beta}}),
\end{equation}
where $\pmb{\mu}_{\pmb{\beta}}, \pmb{\Sigma}_{\pmb{\beta}}$ are defined by \eqref{Hyper:par}. To avoid repetition, BEUGRC denotes the Bayesian estimation of the unrestricted gamma regression coefficients.

Then so as to determine the parameter Bayesian estimations, we rely on the mean squared error loss function. We therefore analyze 100 replicated datasets and for each sample, we estimate $\beta_j$s by obtaining the posterior expected values of MCMC draws after having discarded a burn-in part of the total iterations. 
 We then calculate the mean square error of the estimations using
\begin{equation}
MSE(\hat{\pmb{\beta}})= \dfrac{1}{100} \sum_{k=1}^{100} (\hat{\pmb{\beta}}_{k} - \pmb{\beta}^{true})^T(\hat{\pmb{\beta}}_{k} - \pmb{\beta}^{true}),
\end{equation}
where $\hat{\pmb{\beta}}_{k}$ is the estimation of $\pmb{\beta}$ in the $k$th replication.

{\bf MCMC implementation and results :} We implement the Metropolis-Hastings algorithm by using the \texttt{R} package \textbf{tmvmixnorm} (\cite{Ma}) to set up our simulation study, and produce paths of 15000 length samples for the model parameters. We then discounted 2000 first simulated samples as burn-in to eliminate the effect of initial values. 

The results in Table \ref{MSE:ex} indicate that the proposed Bayesian estimation procedure performs better than other estimators in terms of the mean squared error (MSE). Furthermore, the MSEs of the ridge estimators are significantly lower than the MLE and the MSEs of Bayesian estimators are significantly lower than all classical estimators.
However, all MSEs rise when the level of inter-correlation among the independent variables becomes more intense.
More precisely, when the value of the precision parameter increases from $0.25$ to $0.5$, the MSEs of all estimators (apart from BEGRC) are almost doubled.

Then, Table \ref{bias1} and Table \ref{bias2} display the standard deviation (SD) and the bias of each parameter of the model when $\zeta=0.25$ and $\zeta=0.5$, respectively.
These tables illustrate that the SDs and the bias of the proposed Bayesian estimators of each model parameter are significantly lower than those of other estimators and so the posterior estimates are more accurate than the others.  

\begin{table}[!ht]
  \begin{center}
    \caption{Mean squared errors of the maximum likelihood estimator, two ridge estimators whose parameters are listed in \eqref{ridge.tun}, as well as the posterior expected values in both cases inequality restricted and unrestricted gamma regression models. .}
    \label{MSE:ex}
 \footnotesize
\begin{tabular}{cccccccccc}
    \toprule
        &   &         &&     \multicolumn{3}{c}{Classical}  && \multicolumn{2}{c}{Bayesian}  \\
        \cmidrule{5-7} \cmidrule{9-10}
        &   &         &&     &   \multicolumn{2}{c}{Ridge}  &&    &      \\
              \cmidrule{6-7}
$\zeta$ & $n$ & $\rho$  && \text{MLE} & \text{GRE1}& \text{GRE2} && \text{BEUGRC} & \text{BEGRC}\\
   \hline
\multirow{16}{*}{0.25} & \multirow{4}{*}{25}& 0.80 && 0.16820 & 0.12359 & 0.07577 && 0.06813 & 0.01919 \\
     &   & 0.90 && 0.53948 & 0.19295 & 0.09668 && 0.17301 & 0.02162 \\
     &   & 0.95 && 0.75186 & 0.40213 & 0.05781 && 0.26822 & 0.02211 \\
     &   & 0.99 && 4.45733 & 3.02135 & 0.26769 && 1.58090 & 0.01267 \\
[5pt]
     &  \multirow{4}{*}{50}& 0.80 && 0.14440 & 0.12630 & 0.08465 && 0.04627 & 0.01958 \\
     &   & 0.90 && 0.24093 & 0.12221 & 0.10913 && 0.06925 & 0.02116 \\
     &   & 0.95 && 0.67067 & 0.39471 & 0.25032 && 0.19443 & 0.02925 \\
     &   & 0.99 && 3.55006 & 2.55390 & 0.40920 && 0.95016 & 0.01879 \\
[5pt]
     &  \multirow{4}{*}{100} & 0.80 && 0.10022 & 0.06076  & 0.08074 && 0.02894 & 0.01693 \\
     &     & 0.90 && 0.23761 & 0.10670 & 0.17265 && 0.06555 & 0.02303 \\
     &     & 0.95 && 0.50125 & 0.24363 & 0.30864 && 0.13676 & 0.02824 \\
     &     & 0.99 && 1.95792 & 1.14425 & 0.88240 && 0.53845 & 0.02954 \\
[5pt]
     &  \multirow{4}{*}{200} & 0.80 && 0.08021 & 0.08215 & 0.07094  && 0.02174 & 0.01479 \\
     &     & 0.90 && 0.28022 & 0.20593 & 0.25902 && 0.06800 & 0.02534 \\
     &     & 0.95 && 0.43884 & 0.19984 & 0.34974 && 0.10877 & 0.03149 \\
     &     & 0.99 && 1.49094 & 0.83258 & 1.00776 && 0.40744 & 0.03524 \\
 \hline
\multirow{16}{*}{0.50} & \multirow{4}{*}{25}& 0.80 && 0.28567 & 0.14498 & 0.09961 && 0.12016 & 0.02067 \\
     &   & 0.90 && 0.84251 & 0.34092 & 0.14627 && 0.26354 & 0.02200 \\
     &   & 0.95 && 1.70736 & 0.69756 & 0.23514 && 0.56849 & 0.02745 \\
     &   & 0.99 && 9.45752 & 6.07635 & 0.41636 && 3.40392 & 0.01194 \\
[5pt]
     &  \multirow{4}{*}{50}& 0.80 && 0.22538 & 0.15220  & 0.14384  && 0.07459 & 0.02039 \\
     &   & 0.90 && 0.55065 & 0.20671 & 0.22762 && 0.16047 & 0.02814 \\
     &   & 0.95 && 1.16342 & 0.48282 & 0.37845 && 0.35161 & 0.02776 \\
     &   & 0.99 && 5.46471 & 3.78032 & 1.52912  && 1.64316 & 0.01411 \\
[5pt]
     &  \multirow{4}{*}{100} & 0.80 && 0.19512 & 0.13507 & 0.19248 && 0.05555 & 0.02314 \\
     &     & 0.90 && 0.37146 &  0.19841 & 0.37096  && 0.09896 & 0.02445 \\
     &     & 0.95 && 0.97204 & 0.44306  &  0.58318 && 0.26822 & 0.03047 \\
     &     & 0.99 && 4.49895 & 2.65349  & 2.73289  &&1.18699  & 0.02273 \\
[5pt]
     &  \multirow{4}{*}{200} & 0.80 && 0.13973 & 0.10850  &  0.12686 && 0.03741 & 0.02155 \\
     &     & 0.90 && 0.25639 & 0.09369 & 0.23116 && 0.06984 & 0.02417 \\
     &     & 0.95 && 0.63614 & 0.27421 & 0.54699 && 0.17213 & 0.03168 \\
     &     & 0.99 && 3.81665 & 2.01710 & 2.68077 && 0.95022 & 0.04049 \\
\bottomrule
\end{tabular}
\end{center}
\end{table}

\section{Application to a real dataset}\label{real_app}
The dataset we focus on in this section is a so called ``body fat'' dataset initially studied by \cite{Garcia} and then modeled using the gamma regression model by \cite{Reangsephet} and lately by \cite{Asar}.
The entirety of the dataset may be accessed in \texttt{R} programming language and it is known as TH. data supported by \cite{Hothorn}.
There are 71 observations of healthy female participants included in this dataset, along with nine variables, which are as follows :
age in years (age) $(X_1)$, waist circumference (waistcirc) $(X_2)$, hip circumference (hipcirc) $(X_3)$,
breadth of the elbow (elbowbreadth) $(X_4)$, breadth of the knee (kneebreadth) $(X_5)$, and the sum of logarithms of three anthropometric
measurements that have been divided into four groups (anthro3a $(X_6)$, anthro3b $(X_7)$, anthro3c $(X_8)$, and anthro4 $(X_9)$).

\begin{figure}[hbt!]
\centering
\includegraphics[width=4.5in, height=3.5in]{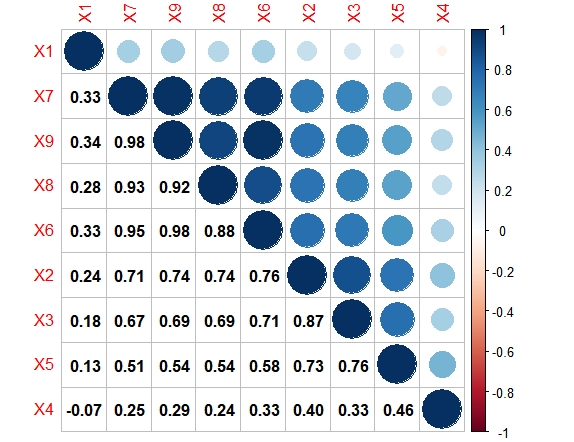}
\caption{Visualization of the correlation matrix.} \label{fig1}
\end{figure}

We start our analyses by conducting an Anderson–Darling (AD) test by using the ad.test function that is included in the gofTest package in \texttt{R}. We used this test to examine whether the gamma regression fits well the response variable (the amount of body fat). 
The AD test statistic value, $0.36082$, 
 confirms that the gamma distribution is a relevant choice for the response variable. 
To investigate the multicollinearity problem through the correlation level between the covariates, a visualization of the correlation matrix  is shown in Figure \ref{fig1} when the corresponding coefficients are reported in the lower tail of the matrix . Visually, there are strong positive relationships between some covariates. Moreover, this conclusion is confirmed by the condition number of the matrix $X^T \hat{\mathcal{C}} X$ that is equal to $4026.235$. We recall here that the condition number of a matrix is the square root of the ratio of the highest eigenvalue to the minimum one.

Regarding the form of the restriction matrix, we focus on the studies and results of \cite{Reangsephet} and \cite{Asar}  and we set the following restrictions on the model parameters
\begin{equation} \label{real.rest}
  \beta_6 \leq0, \beta_7\geq 0, \beta_8\geq, \beta_9\geq0.
\end{equation}

We consider the quadratic loss function and compute the posterior expected values for the parameter restricted gamma regression model as well as for the gamma regression model without taking into account any restriction. The total number of the MCMC iterations is 15000 MCMC with 2000 first iterations as burn-in.
Table \ref{RES.REAL} displays that despite the similarity of the estimate values, the related standard errors for the Bayesian cases are  nevertheless substantially lower than the classical cases.

\begin{table}[!ht]
  \begin{center}
    \caption{For {\bf  body fat dataset :} (right side) Posterior estimates obtained for restricted and unrestricted gamma regression models. (left side) MLE indicates the maximum likelihood estimator and GRE is obtained by using the ridge regression model.  And the standard errors associated with each set of the estimations are displayed in the brackets. }
    \label{RES.REAL}
 \footnotesize
\begin{tabular}{lcccccc}
  \toprule
             &&\multicolumn{2}{c}{Classical}  && \multicolumn{2}{c}{Bayesian}  \\
        \cmidrule{3-4} \cmidrule{6-7}
              && \text{MLE}   & \text{GRE}&& \text{BEUGRC} & \text{BEGRC} \\
  \toprule
  Intercecpt  &&-0.2195 (0.8154)&-0.1012 (0.4289)&& -0.2219 (0.1686)& -0.2492 (0.1592)\\
  Age         &&0.0016 (0.0033)&0.0016 (0.0032)&&  0.0016 (0.0007)&  0.0016 (0.0007)\\
  Waistcirc   &&0.0042 (0.0060)&0.0045 (0.0055)&&  0.0041 (0.0015)&  0.0042 (0.0014)\\
  Hipcirc     &&0.0108 (0.0071)&0.0105 (0.0068)&&  0.0106 (0.0018)&  0.0107 (0.0017)\\
  Elbowbreadth&&0.0141 (0.1037)&0.0063 (0.0828)&&  0.0146 (0.0237)&  0.0172 (0.0227)\\
  Kneebreadth &&0.0426 (0.0620)&0.0412 (0.0598)&&  0.0445 (0.0148)&  0.0441 (0.0171)\\
  Anthro3a    &&-0.1227 (0.5417)&-0.0580 (0.3217)&& -0.1275 (0.1166)& -0.1636 (0.0850)\\
  Anthro3b    &&0.1404 (0.5474)&0.1309 (0.3353&&  0.1450 (0.1250)&  0.1459 (0.0822)\\
  Anthro3c    &&0.1292 (0.2094)&0.1304 (0.1879)&&  0.1278 (0.0463)&  0.1240 (0.0431)\\
  Anthro4     &&0.1637 (0.6547)&0.1140 (0.3288)&&  0.1663 (0.1436)&  0.1938 (0.0946)\\
  \bottomrule
\end{tabular}
\end{center}
\end{table}

\section{Conclusion}\label{conclusion}
Our focus is on the parameter estimating challenge for the gamma regression model, when the design matrix suffers multicollinearity issue and the model parameters have some linear inequality restrictions.The restrictions can be available from a priori knowledge and are essential to be taken into account so as to preserve the coherence of the model structure.
We therefore present in this paper, a Bayesian procedure for estimating the restricted parameters of the gamma regression model with log link function. We note that \cite{Ghosal} have provided a Bayesian framework for the generalized linear models, however, their methodology is not satisfactory for the gamma regression models with the log link function.
We perform a simulation study to check the accuracy of the estimation procedure and the results show that the suggested method offers more accurate parameter estimation compared to that provided by well-known estimators (MLE and Ridge regression model).
We then apply the method to analyze body fat dataset and the results show that the Bayesian method introduced in Section \ref{sec3} results in parameter estimations whose standard errors are considerably less than other classical methods.


\newpage
\section{Appendix}

\begin{table}[!ht]
  \begin{center}
    \caption{Standard deviations and bias of the estimators when $\zeta=0.25$.} \label{bias1}
    \label{total1}
 \footnotesize
\begin{tabular}{ccccccccc}
    \toprule
 $\rho$ &  & \text{MLE} & \text{GRE1}& \text{GRE2} & \text{BEUGRC} & \text{BEGRC}\\
   \hline
   $n=25$ & &      &   &   &   &   &   \\
   \hline
 \multirow{4}{*}{0.80} & $\beta_1$ & 0.4240(0.0260) & 0.3380(-0.0579) & 0.2697(-0.0621) & 0.2814(0.0204) &0.1499(0.0270) \\
& $\beta_2$ & 0.4232(-0.0590) &0.3606(-0.1306) & 0.2577(-0.0822) & 0.2642(-0.0418) & 0.1231(-0.0135)\\
& $\beta_3$ & 0.3478(-0.0393)& 0.2969(-0.0992) & 0.2404(-0.1014) &0.2178 (-0.0315) & 0.1207(-0.0140)\\
& $\beta_4$ & 0.4400(0.0190)& 0.3572(-0.0933) & 0.2805(-0.0964) & 0.2749(0.0049) & 0.1548(0.0166)\\
  [5pt]
\multirow{4}{*}{0.90} & $\beta_1$ & 0.7612(0.0062) & 0.4887(-0.0303) & 0.3326(-0.0834) & 0.4185(-0.0059) &0.1544(0.0052) \\
& $\beta_2$ & 0.8206(-0.0988) & 0.4409(-0.0334) & 0.2581(-0.0610) & 0.4536(-0.0434) &0.1614(0.0002) \\
& $\beta_3$ & 0.6463(-0.0361) & 0.4089(-0.0472) & 0.3094(-0.0636) & 0.3719(-0.0318) &0.1319(-0.0114) \\
& $\beta_4$ & 0.7028(0.0641) & 0.4144(-0.0565) & 0.3148(-0.0603)  & 0.4196(0.0281) &0.1409(0.0075) \\
  [5pt]
\multirow{4}{*}{0.95} & $\beta_1$ & 0.8532(0.1286) & 0.6594(0.0558) & 0.2580(-0.0607) & 0.4851(0.0961) & 0.1472(0.0246) \\
& $\beta_2$ & 0.8485(-0.0091) & 0.6131(0.0279)  & 0.2207(-0.0348) & 0.5246(0.0181) & 0.1497(0.0045) \\
& $\beta_3$ & 0.7112(-0.1107) & 0.4669(-0.1386) & 0.2127(-0.0546) & 0.4412(-0.0866) & 0.1231(-0.0168) \\
& $\beta_4$ & 1.0260(-0.0491) & 0.7538(-0.0680) & 0.2416(-0.0814) & 0.5972(-0.0674) & 0.1704(0.0113) \\
  [5pt]
\multirow{4}{*}{0.99} & $\beta_1$ & 1.8935(0.0493) & 1.5741(0.0540) & 0.5514(0.0090) & 1.1061(0.0536) & 0.1113(0.0053) \\
& $\beta_2$ & 2.2292(-0.1157) & 1.7512(-0.0645) & 0.5351(-0.0899) & 1.2973(0.0090) & 0.1221(0.0024) \\
& $\beta_3$ & 2.3378(-0.2697) & 1.9263(-0.2942) & 0.3398(-0.1038) & 1.4165(-0.1890) & 0.1117(-0.0039) \\
& $\beta_4$ & 1.9541(0.2846)  & 1.6728(0.2432) & 0.5964(-0.0296) & 1.1952(0.0835) & 0.1064(0.0052) \\
   \hline
   $n=50$ & &      &   &   &   &   &   \\
   \hline
 \multirow{4}{*}{0.80} & $\beta_1$ & 0.3439(-0.0561) & 0.3144(-0.0865) & 0.2931(-0.0507) & 0.2032(-0.0361) & 0.1236(-0.0187) \\
& $\beta_2$ &  0.3251(0.0123)  & 0.3302(-0.0765) & 0.2501(-0.0041) & 0.1781(0.0020) & 0.1342(0.0011) \\
& $\beta_3$ &  0.4400(-0.0326) & 0.3624(-0.1257) & 0.2798(-0.0684) & 0.2389(-0.0223) & 0.1560(-0.0065) \\
& $\beta_4$ &  0.4019(-0.0049) & 0.3635(-0.0973) & 0.3256(-0.0442) & 0.2345(0.0045) & 0.1452(0.0054) \\
  [5pt]
\multirow{4}{*}{0.90} & $\beta_1$ &  0.4918(0.0167) & 0.3747(-0.0260) & 0.3185(-0.0423) & 0.2625(-0.0105) & 0.1511(-0.0015) \\
& $\beta_2$ &  0.4850(-0.0029) & 0.3107(-0.0097) & 0.3273(0.0115) & 0.2637(0.0065) & 0.1454(0.0020) \\
& $\beta_3$ &  0.5228(-0.0105) & 0.3958(-0.0403) & 0.3711(-0.0283) & 0.2739(-0.0071) & 0.1509(0.0005) \\
& $\beta_4$ &  0.4679(-0.0607) & 0.3074(-0.0565) & 0.2995(-0.0464) & 0.2558(-0.0263) & 0.1367(-0.0097) \\
  [5pt]
\multirow{4}{*}{0.95} & $\beta_1$ &  0.7790(0.0678) & 0.5763(0.0224) & 0.4002(-0.0133) & 0.4101(0.0296) & 0.1727(0.0105) \\
     & $\beta_2$ &  0.7976(-0.1855) & 0.5844(-0.1932) & 0.4720(-0.0970) & 0.4344(-0.1112) & 0.1374(-0.0300) \\
     & $\beta_3$ &  0.8678(-0.0954) & 0.7014(-0.0418) & 0.5902(-0.0377) & 0.4605(-0.0458) & 0.1735(-0.0102) \\
     & $\beta_4$ &  0.7980(0.1677) & 0.6123(0.1190) & 0.5151(0.0598) & 0.4369(0.1006) & 0.1913(0.0405) \\
  [5pt]
\multirow{4}{*}{0.99} & $\beta_1$ &  1.9689(-0.4562) & 1.7291(-0.4212) & 0.6707(-0.1546) & 1.0127-0.2691() & 0.1183(-0.0307) \\
      & $\beta_2$ &  1.8492(0.3026) & 1.5382(0.2455) & 0.6321(0.0930) & 0.9749(0.1307) & 0.1493(0.0168) \\
      & $\beta_3$ &  1.7502(0.0721) & 1.4732(0.0906) & 0.5178(-0.0379) & 0.9175(0.0300) & 0.1286(-0.0021) \\
     & $\beta_4$ &  1.9173(0.0183) & 1.5949(-0.0109) & 0.7075(0.0301) & 0.9616(0.0707) & 0.1473( 0.0133) \\
\hline
\end{tabular}
\end{center}
\end{table}

\begin{table}[!ht]
  \begin{center}
    \caption*{\textbf{Table 3} Continued.}
    \label{total2}
 \footnotesize
\begin{tabular}{ccccccccc}
    \toprule
 $\rho$ &  & \text{MLE} & \text{GRE1}& \text{GRE2} & \text{BEUGRC} & \text{BEGRC}\\
   \hline
   $n=100$ & &      &   &   &   &   &   \\
   \hline
 \multirow{4}{*}{0.80} & $\beta_1$ & 0.3081(-0.0455) & 0.2433(-0.0837) & 0.2778(-0.0526) & 0.1741(-0.0213) & 0.1255(-0.0134) \\
      & $\beta_2$ & 0.3523(0.0090) & 0.2467(-0.0315) & 0.3173(0.0013) & 0.1851(0.0070) & 0.1472(0.0123) \\
      & $\beta_3$ & 0.3100(-0.0489)& 0.2362(-0.0919) & 0.2664(-0.0399) & 0.1656(-0.0309) & 0.1209(-0.0215) \\
      & $\beta_4$ & 0.2898(0.0345) & 0.2293(-0.0214) & 0.2695(0.0181) & 0.1523(0.0157) & 0.1245(0.0081) \\
  [5pt]
\multirow{4}{*}{0.90} & $\beta_1$ & 0.4516(-0.0289) & 0.3037(-0.0762) & 0.3917(-0.0183) & 0.2467(-0.0148) & 0.1399(-0.0111) \\
      & $\beta_2$ & 0.4544(-0.0872) & 0.3028(-0.1148) & 0.3721(-0.1012) & 0.2371(-0.0489) & 0.1419(-0.0241) \\
      & $\beta_3$ & 0.4934(0.0227) & 0.3233(-0.0071) & 0.4375(0.0106) & 0.2510(0.0102) & 0.1487(0.0059) \\
      & $\beta_4$ & 0.5444(0.0260) & 0.3473(-0.0533) & 0.4504(0.0240) & 0.2863(0.0101) & 0.1745(0.0067) \\
  [5pt]
 \multirow{4}{*}{0.95} & $\beta_1$ & 0.7341(-0.1217) & 0.5351(-0.0961) & 0.5978(-0.0974) & 0.3855(-0.0534) & 0.1764(-0.0087) \\
      & $\beta_2$ & 0.6575(-0.0332) & 0.4609(-0.0980) & 0.5523(-0.0423) & 0.3399(-0.0278) & 0.1470(-0.0227) \\
      & $\beta_3$ & 0.7246(0.1242) & 0.4939(0.0843) & 0.5500(0.0962) & 0.3873(0.0520) & 0.1728(0.0203) \\
      & $\beta_4$ & 0.7050(-0.0152) & 0.4641(-0.0092) & 0.5112(-0.0075) & 0.3634(0.0010) & 0.1746(0.0051) \\
  [5pt]
 \multirow{4}{*}{0.99} & $\beta_1$ & 1.5133(-0.2864) & 1.1771(-0.1824) & 1.0937(-0.1604) & 0.7895(-0.1470) & 0.1527(-0.0257) \\
      & $\beta_2$ & 1.2649(0.0437) &0.9227(-0.0224) & 0.8389(-0.0058) & 0.6859(0.0372) & 0.1789(0.0107) \\
      & $\beta_3$ & 1.3716(0.2987) &1.0101(0.2570) & 0.9462(0.1609) & 0.7008(0.1549) & 0.1834(0.0258) \\
      & $\beta_4$ & 1.3980(-0.0990) &1.1211(-0.0923) & 0.8463(-0.0403) & 0.7338(-0.0684) & 0.1695(-0.0146) \\
    \hline
   $n=200$ & &      &   &   &   &   &   \\
   \hline
 \multirow{4}{*}{0.80} & $\beta_1$ & 0.2720(-0.0399) & 0.2740(-0.0719) & 0.2520(-0.0214) & 0.1448(-0.0223) & 0.1150(-0.0164) \\
      & $\beta_2$ & 0.2882(-0.0251) & 0.2737(-0.1035) & 0.2735(-0.0280) & 0.1475(-0.0163) & 0.1231(-0.0138) \\
      & $\beta_3$ & 0.2879(0.0546)  & 0.2904(-0.0393) & 0.2711(0.0414) & 0.1507(0.0277) & 0.1315(0.0258) \\
      & $\beta_4$ & 0.2800(-0.0218) & 0.9021(-0.0979) & 0.2661(-0.0319) & 0.1442(-0.0087) & 0.1135(-0.0079) \\
  [5pt]
\multirow{4}{*}{0.90} & $\beta_1$ &  0.6472(-0.0497) & 0.5682(-0.1092) & 0.6385(-0.0528) & 0.3034(-0.0217) & 0.1627(0.0042) \\
      & $\beta_2$ &  0.6279(0.0992) & 0.5648(0.0263) & 0.5969(0.0889) & 0.3119(0.0382) & 0.1842(0.0139) \\
      & $\beta_3$ &  0.3903(-0.0062) & 0.2767(-0.0503) & 0.3601(-0.0096) & 0.2079(0.0008) & 0.1515(0.0002) \\
      & $\beta_4$ &  0.3811(-0.0946) & 0.2912(-0.1160) & 0.3678(-0.0822) & 0.1948(-0.0476) & 0.1327(-0.0348) \\
  [5pt]
\multirow{4}{*}{0.95} & $\beta_1$ &  0.5890(-0.0801) & 0.4192(-0.0574) & 0.5890(-0.0801) & 0.3005(-0.0495) & 0.1626(-0.0263) \\
      & $\beta_2$ &  0.6746(-0.0177) & 0.4681(-0.0194) & 0.6746(-0.0177) & 0.3483(0.0037) & 0.1934(0.0096) \\
      & $\beta_3$ &  0.6933(0.0345) & 0.4590(-0.0013) & 0.6933(0.0345) & 0.3367(0.0124) & 0.1737(0.0013) \\
      & $\beta_4$ &  0.6943(0.0166) & 0.4451(-0.0057) & 0.6943(0.0166) & 0.3344(0.0026) & 0.1802(-0.0015) \\
  [5pt]
\multirow{4}{*}{0.99} & $\beta_1$ &  0.9455(0.0124) & 0.7058(-0.0704) & 0.7678(0.0050) & 0.5104(-0.0095) & 0.1601(-0.0192) \\
      & $\beta_2$ &  1.3014(-0.0513)& 1.0379(-0.0250) & 1.1599(0.0089) & 0.6890(-0.0093) & 0.2012(0.0013) \\
      & $\beta_3$ &  1.2750(0.0958) & 0.9545(0.0255) & 0.9897(0.0065) & 0.6656(0.0482) & 0.1914(0.0067) \\
     & $\beta_4$ &  1.3373(-0.1021)& 0.9329(-0.0280) & 1.0738(-0.0636) & 0.6802(-0.0529) & 0.1980(-0.0013) \\
\hline
\end{tabular}
\end{center}
\end{table}

\begin{table}[!ht]
  \begin{center}
    \caption{Standard deviations and bias of the estimators when $\zeta=0.5$.}\label{bias2}
    \label{total3}
 \footnotesize
\begin{tabular}{cccccccc}
    \toprule
 $\rho$ &  & \text{MLE} & \text{GRE1}& \text{GRE2} & \text{BEUGRC} & \text{BEGRC}\\
    \hline
   $n=25$ & &      &   &   &   &   &   \\
   \hline
 \multirow{4}{*}{0.80} & $\beta_1$ & 0.6162(-0.0768) & 0.3921(-0.1010) & 0.3152(-0.0904) & 0.4050(-0.0569) & 0.1633(0.0195) \\
      & $\beta_2$ & 0.4956(-0.0378) & 0.3272(-0.1582) & 0.2813(-0.0752) & 0.3005(-0.0191) & 0.1268(-0.0044) \\
      & $\beta_3$ & 0.4663(-0.0103) & 0.3879(-0.0895) & 0.3311(-0.0802) & 0.3084(-0.0185) & 0.1301(0.0094) \\
      & $\beta_4$ & 0.5502(0.0333) & 0.3491(-0.0938) & 0.3026(-0.0477) & 0.3634(0.0085) & 0.1512(0.0216) \\
  [5pt]
 \multirow{4}{*}{0.90} & $\beta_1$ & 0.8566(0.0249) & 0.5633(-0.0634) & 0.3472(0.0155) & 0.4856(-0.0198) &0.1423(0.0079) \\
      & $\beta_2$ & 0.8888(-0.1142) & 0.5138(-0.0489) & 0.3929(-0.1050) & 0.5133(-0.0472) &0.1409(-0.0050) \\
      & $\beta_3$ & 0.8760(-0.0090) & 0.5194(-0.1094) & 0.3030(-0.0577) & 0.5064(-0.0309) &0.1434(0.0013) \\
      & $\beta_4$ & 1.0478(-0.0360) & 0.7013(-0.1256) & 0.4454(-0.1052) & 0.5523(-0.0208) &0.1671(0.0145) \\
  [5pt]
\multirow{4}{*}{0.95} & $\beta_1$ & 1.3398(0.0386) & 0.7904(0.0309) & 0.4130(-0.0112) & 0.7722(0.0499) & 0.1795(0.0106) \\
      & $\beta_2$ & 1.3618(0.0782) & 0.8997(0.0203) & 0.5043(0.0060) & 0.7785(0.0106) & 0.1807(0.0217) \\
      & $\beta_3$ & 1.2626(0.0387) & 0.8375(-0.0088) & 0.4566(-0.0211) & 0.7162(-0.0133) & 0.1600(0.0097) \\
      & $\beta_4$ & 1.2586(-0.2466) & 0.8039(-0.1861) & 0.5417(-0.1496) & 0.7497(-0.1291) & 0.1404(-0.0062) \\
  [5pt]
 \multirow{4}{*}{0.99} & $\beta_1$ & 3.0568(0.0041) & 2.3807(0.0221) & 0.3452(-0.0239) & 1.7549(-0.0065) &0.1009(0.0082) \\
      & $\beta_2$ & 3.4152(0.4109) & 2.7856(0.2949) & 0.7529(0.1748) & 2.1606(0.2459) & 0.1317(0.0232) \\
      & $\beta_3$ & 2.7757(-0.0818) & 2.0698(-0.0099) & 0.6773(-0.0902) & 1.6430(-0.1044) & 0.1039(0.0092) \\
      & $\beta_4$ & 3.0260(-0.4057) & 2.5674(-0.3979) & 0.6920(-0.1373) & 1.7862(-0.2064) &0.0961(-0.0022) \\
     \hline
   $n=50$ & &      &   &   &   &   &   \\
   \hline
 \multirow{4}{*}{0.80} & $\beta_1$ & 0.4675(-0.0274) & 0.3925(-0.1040) & 0.3541(-0.0102) & 0.2738(-0.0305) & 0.1383(-0.0071) \\
      & $\beta_2$ &  0.4940(-0.0299) & 0.3640(-0.1099) & 0.4058(-0.0510) & 0.2833(-0.0182) & 0.1412(0.0067) \\
      & $\beta_3$ &  0.4805(0.0411) & 0.3860(-0.0850) & 0.3874(-0.0324) & 0.2778(0.0173) & 0.1555(0.0132) \\
      & $\beta_4$ &  0.4548(-0.0831) & 0.3591(-0.1411) & 0.3654(-0.0603) & 0.2580(-0.0289) & 0.1373(-0.0039) \\
  [5pt]
\multirow{4}{*}{0.90} & $\beta_1$ &  0.7081(-0.0065) & 0.4121(-0.0218) & 0.4764(0.0177) & 0.3844(-0.0049) & 0.1588(0.0020) \\
     & $\beta_2$ &  0.8043(-0.0377) & 0.4906(0.0048) & 0.5011(-0.0427) & 0.4359(-0.0221) & 0.1874(0.0027) \\
     & $\beta_3$ &  0.7232(-0.0010) & 0.4494(-0.0544) & 0.4682(-0.0564) & 0.3847(-0.0116) & 0.1628(0.0007) \\
     & $\beta_4$ &  0.7426(-0.0220) & 0.4660(-0.0445) & 0.4652(-0.0247) & 0.4021(-0.0135) & 0.1636(0.0090) \\
  [5pt]
\multirow{4}{*}{0.95} & $\beta_1$ &  1.0917(0.0182) & 0.7555(-0.0666) & 0.6747(-0.0097) & 0.5951(-0.0011) & 0.1690(0.0062) \\
     & $\beta_2$ &  1.0518(0.0350) & 0.6358(-0.0113) & 0.6346(-0.0187) & 0.5963(0.0214) & 0.1644(0.0017) \\
     & $\beta_3$ &  1.0152(-0.0357) & 0.6891(-0.0027) & 0.5650(-0.0060) & 0.5715(-0.0201) & 0.1587(-0.0094) \\
     & $\beta_4$ &  1.1642(-0.1167) & 0.6972(-0.1012) & 0.5878(-0.0767) & 0.6147(-0.0737) & 0.1768(-0.0009) \\
  [5pt]
\multirow{4}{*}{0.99} & $\beta_1$ &  2.5045(0.3945) & 2.1119(0.4388) & 1.1324(0.1975) & 1.3598(0.2593) & 0.1299(0.0142) \\
     & $\beta_2$ &  2.3037(-0.3614) & 1.8684(-0.3925) & 1.4361(-0.2283) & 1.2458(-0.1821) & 0.1006(-0.0214) \\
     & $\beta_3$ &  2.2386(0.0835) & 1.7983(0.0998) & 1.1118(-0.0661) & 1.2033(0.0099) & 0.1189(-0.0101) \\
     & $\beta_4$ &  2.2639(-0.2577) & 1.9098(-0.2842) & 1.2244(-0.0445) & 1.2873(-0.1764) & 0.1214(-0.0189) \\
\hline
\end{tabular}
\end{center}
\end{table}

\begin{table}[!ht]
  \begin{center}
  \caption*{\textbf{Table 4} Continued.}
    \label{total4}
 \footnotesize
\begin{tabular}{ccccccccc}
    \toprule
 $\rho$ &  & \text{MLE} & \text{GRE1}& \text{GRE2} & \text{BEUGRC} & \text{BEGRC}\\
      \hline
   $n=100$ & &      &   &   &   &   &   \\
   \hline
 \multirow{4}{*}{0.80} & $\beta_1$ & 0.4534(0.0109) & 0.3562(-0.1053) & 0.3710(0.0039) & 0.2406(0.0055) & 0.1581(0.0092) \\
      & $\beta_2$ & 0.4192(-0.0570) & 0.3416(-0.1311) & 0.3622(-0.0618) & 0.2244(-0.0308) & 0.1442(-0.0167) \\
      & $\beta_3$ & 0.4452(-0.0078) & 0.3342(-0.0982) & 0.3877(0.0004) & 0.2333(0.0029) & 0.1516(0.0086) \\
      & $\beta_4$ & 0.4533(0.0004) & 0.3732(-0.1129) & 0.3739(-0.0135) & 0.2464(-0.0085) & 0.1558(0.0068) \\
  [5pt]
\multirow{4}{*}{0.90} & $\beta_1$ & 0.5808(-0.0080) & 0.4395(-0.0735) & 0.4706(-0.0055) & 0.2969(-0.0165) & 0.1416(-0.0080) \\
      & $\beta_2$ & 0.6489(-0.0760) & 0.4691(-0.0978) & 0.5611(-0.0322) & 0.3271(-0.0238) & 0.1538(-0.0077) \\
      & $\beta_3$ & 0.6078(-0.0277) & 0.4152( -0.0792) & 0.4941(-0.0555) & 0.3219(-0.0039) & 0.1576(0.0075) \\
      & $\beta_4$ & 0.6038(0.0432) & 0.4410(-0.0061) & 0.5014(0.0161) & 0.3165(0.0032) & 0.1734(0.0034) \\
  [5pt]
 \multirow{4}{*}{0.95} & $\beta_1$ & 0.9179(0.0643) & 0.5773(-0.0174) & 0.7603(0.0324) & 0.4743(0.0373) & 0.1677(-0.0010) \\
      & $\beta_2$ & 0.8740(-0.1553) & 0.8740(-0.1553) & 0.5885(-0.1116) & 0.4955(-0.1144) & 0.1590(-0.0397) \\
      & $\beta_3$ & 0.9325(-0.0437) & 0.9325(-0.0437) & 0.6729(-0.0813) & 0.4795(-0.0146) &0.1675(-0.0058) \\
      & $\beta_4$ & 1.1919(0.0178) & 1.1919(0.0178) &  0.7988(0.0058) & 0.6068(0.0147) &0.1994(0.0166) \\
  [5pt]
 \multirow{4}{*}{0.99} &$\beta_1$ & 2.0674(0.1807) & 1.4760(0.1131) & 1.5522(0.0958) & 1.1042(0.0840) & 0.1610(0.0132) \\
      & $\beta_2$ & 2.0577(-0.2009) & 1.6722(-0.2440) & 1.5500(-0.1564) & 1.0924(-0.1177) & 0.1409(-0.0183) \\
      & $\beta_3$ & 2.2970(0.1105) & 1.7073(0.1576) & 1.8949(0.0860) & 1.1571(0.0198) & 0.1613(-0.0054) \\
      & $\beta_4$ & 2.0638(-0.2178) & 1.6461(-0.1529) & 1.6068(-0.1283) & 1.0099(-0.0560) & 0.1379(-0.0220) \\
     \hline
   $n=200$ & &      &   &   &   &   &   \\
   \hline
 \multirow{4}{*}{0.80} & $\beta_1$ & 0.3666(0.0166) & 0.3172(-0.1067) & 0.3389(0.0129) & 0.1920(0.0032) & 0.1502(0.0011) \\
      & $\beta_2$ &  0.3877(-0.0606) & 0.3026(-0.1383) & 0.3763(-0.0614) & 0.1988(-0.0354) & 0.1398(-0.0191) \\
      & $\beta_3$ &  0.3590(0.0403) & 0.3072(-0.0719) & 0.3348(0.0355) & 0.1859(0.0243) & 0.1524(0.0172) \\
     & $\beta_4$ &  0.3785(-0.0455) & 0.3050(-0.1502) & 0.3701(-0.0419) & 0.1943(-0.0229) & 0.1446(-0.0123) \\
  [5pt]
 \multirow{4}{*}{0.90} & $\beta_1$ &  0.5063(0.0726) & 0.2732(-0.0288) & 0.4733(0.0766) & 0.2589(0.0460) & 0.1688(0.0315) \\
      & $\beta_2$ &  0.4692(-0.0431) & 0.2786(-0.0634) & 0.4353(-0.0292) & 0.2474(-0.0286) & 0.1464(-0.0239) \\
      & $\beta_3$ &  0.5709(-0.0592) & 0.3710(-0.0410) & 0.5505(-0.0622) & 0.2931(-0.0276) & 0.1605(-0.0042) \\
      & $\beta_4$ &  0.4704(-0.0385) & 0.2859(-0.0189) & 0.4513(-0.0555) & 0.2517(-0.0295) & 0.1410(-0.0227) \\
  [5pt]
\multirow{4}{*}{0.95} & $\beta_1$ &  0.7544(0.0984) & 0.5177(0.0139) & 0.7129(0.0878) & 0.3965(0.0578) & 0.1804(0.0142) \\
      & $\beta_2$ &  0.8485(-0.1131) & 0.5725(-0.1116) & 0.7771(-0.0792) & 0.4358(-0.0689) & 0.1698(-0.0238) \\
      & $\beta_3$ &  0.8052(-0.0107) & 0.5113(-0.0249) & 0.7645(-0.0201) & 0.4196(-0.0047) & 0.1887(-0.0026) \\
      & $\beta_4$ &  0.7797(-0.0471) & 0.4871(-0.0097) & 0.7035(-0.0627) & 0.4042(-0.0265) & 0.1737(-0.0074) \\
 [5pt]
 \multirow{4}{*}{0.99} & $\beta_1$ &  2.0441(0.3429) & 1.5988(0.2453) & 1.7392(0.2861) & 1.0201(0.1530) & 0.2206(0.0236) \\
      & $\beta_2$ &  1.8849(-0.2827) & 1.3445(-0.1107) & 1.6236(-0.1702) & 0.9659(-0.1469) & 0.1978(-0.0214) \\
      & $\beta_3$ &  2.0654(0.0266) & 1.5035(-0.0951) & 1.6870(-0.0959) & 0.9984(0.0438) & 0.1990(0.0073) \\
      & $\beta_4$ &  1.7893(-0.1472) & 1.1965(-0.1066) & 1.4823(-0.0777) & 0.9025(-0.0846) & 0.1859(-0.0230) \\
\hline
\end{tabular}
\end{center}
\end{table}


\end{document}